\documentclass{article}

\usepackage[final]{neurips_ts4h_2022}

\usepackage[utf8]{inputenc} 
\usepackage[T1]{fontenc}    
\usepackage{hyperref}       
\usepackage{url}            
\usepackage{booktabs}       
\usepackage{amsfonts}       
\usepackage{nicefrac}       
\usepackage{microtype}      
\usepackage{xcolor}         
\usepackage{graphicx}
\usepackage{amsmath}

\title{MAEEG: Masked Auto-encoder for\\EEG Representation Learning}

\author{%
    Hsiang-Yun Sherry Chien, 
    Hanlin Goh,
    Christopher M.~Sandino,
    Joseph Y.~Cheng \\
    Apple\\
  \texttt{\{sherry.chien,hanlin,csandino,jycheng\}@apple.com} \\
}

\begin{document}

\maketitle

\begin{abstract}
  Decoding information from bio-signals such as EEG, using machine learning has been a challenge due to the small data-sets and difficulty to obtain labels. We propose a reconstruction-based self-supervised learning model, the masked auto-encoder for EEG (MAEEG), for learning EEG representations by learning to reconstruct the masked EEG features using a transformer architecture. We found that MAEEG can learn representations that significantly improve sleep stage classification ($\sim$ 5\% accuracy increase) when only a small number of labels are given. We also found that input sample lengths and different ways of masking during reconstruction-based SSL pretraining have a huge effect on downstream model performance. Specifically, learning to reconstruct a larger proportion and more concentrated masked signal results in better performance on sleep classification. Our findings provide insight into how reconstruction-based SSL could help representation learning for EEG. 
\end{abstract}


\section{Introduction}
Decoding information from Electroencephalography (EEG) and other bio-signal modalities has enabled health-related clinical applications, such as sleep pattern detection \citep{ghassemi2018you}. Recently, machine learning models have been effectively applied on classifying EEG signals, mostly via supervised learning \citep{lawhern2018eegnet}. However, such methods rely on a large number of labels for bio-signal data, which are usually difficult and expensive to obtain. One way to improve accuracy when decoding EEG signals using only a small number of labels is to learn useful representations from unlabeled data via self-supervised learning (SSL). 

SSL has been widely explored in the field of computer vision and speech processing. The idea of SSL is to boost the performance of a downstream task by learning meaningful representations through an SSL task. One way to learn representations is through masking and reconstructing the input signals (i.e., reconstruction-based SSL). For example, the representations of speech learned from the Wav2Vec series of SSL models significantly improve the performance of various downstream tasks, such as automatic speech recognition and speech emotion recognition \citep{baevski2020wav2vec, mitra2022speech}. Given the success of reconstruction-based SSL methods in other domains, it is desirable to learn whether and how such methods could be adapted to physiological time-series data, such as EEG.

In this study, we explore representation learning using reconstruction-based SSL on EEG data. We propose an SSL model, masked auto-encoder for learning EEG representations (MAEEG), which can learn EEG representations by reconstructing the raw signal from masked features. We found that MAEEG pretraining learns meaningful EEG representations, which yield better performance on sleep stage classification. We further explore how masking may affect SSL and downstream task performance, and found that in general, a higher probability and more concentrated masking yields better task performance.

\section{Background and Related Work}
SSL methods for time-series data have been widely applied in domains such as natural language and speech processing. Recently, studies have also explored various SSL methods for bio-signal data, such as EEG. For example, combining various ways of EEG data augmentation and contrastive learning, \citet{mohsenvand2020contrastive} showed improved performance on several EEG classification tasks. On the other hand, \citet{banville2021uncovering} proposed three SSL methods aiming to learn representations by discriminating the temporal relationships between the EEG samples. These studies showed that a SSL-pretrained encoder could produce features that outperformed supervised learning when there is only a small number of labeled data. 

Another type of SSL methods for time-series data learn representations through masking and reconstructing features, and we refer to them as reconstruction-based SSL in the current study. Reconstruction-based SSL usually contains several stages: (1) The raw input signal is first encoded as features using a convolutional encoder; (2) part of the features are ``masked" by setting the features to a certain value; (3) the masked features are sent to a second encoder, usually a transformer encoder, which aims to reconstruct the masked part of the features using the unmasked features as context; (4) finally, contrastive loss, or reconstruction loss is calculated to optimize the SSL task. Inspired by the Wav2Vec2 model, \citet{kostas2021bendr} proposed BENDR, a reconstruction-based SSL with contrastive learning for processing EEG signals. While the idea of using massive data (i.e.,\ the TUH dataset, \citep{obeid2016temple}) to learn EEG representations using BENDR seems to improve some downstream tasks, the results are not consistent across different data-sets/tasks. This could be due to the inconsistency of data-sets and model settings between pretraining and downstream stages. Overall, it remains unclear if their reconstruction-based SSL methods could learn useful representations for EEG signals. 

In this study, we examine whether reconstruction-based SSL methods are effective for learning EEG representations, and how different ways of masking would affect representation learning. Specifically, we propose an SSL model, MAEEG, inspired by both BENDR and the MAE model proposed in computer vision \citep{he2022masked}, to compare with BENDR and to broaden our understanding of reconstruction-based SSL for EEG. 

\section{Method}
\subsection{Sleep EEG Data-set}
To examine how reconstruction-based SSL may help learn useful representations for classifying sleep stages from EEG signals, we use the data-set provided from 2018 PhysioNet Challenge \citep{ghassemi2018you}.The data-set contains overnight polysomnography of 994 subjects (i.e.,\ around 7,000 hours of data, 65\% male, mean age: 55y$\pm$14.4), which was split into a training set (596 subjects), a validation set (198 subjects) and a testing set (200 subjects) as in \citet{ghassemi2018you}. Each polysomnogram contains simultaneous recordings of 6-channel EEG, ECG and respiratory signals during sleep and two sets of labels: 5 sleep stages and respiratory-effort related arousals. The labels were assigned by a trained sleep technician for each 30-sec non-overlapping window. Here, we focused on using the 6-channel EEG signals to classify 5 stages of sleeping (i.e.,\ wake, N1, N2, N3, REM). All the models and experiments presented in this section are built and run on PyTorch 1.8.1.
\subsection{Reconstruction-based SSL models}
\begin{figure*}[ht]
\centerline{\includegraphics[width=\columnwidth]{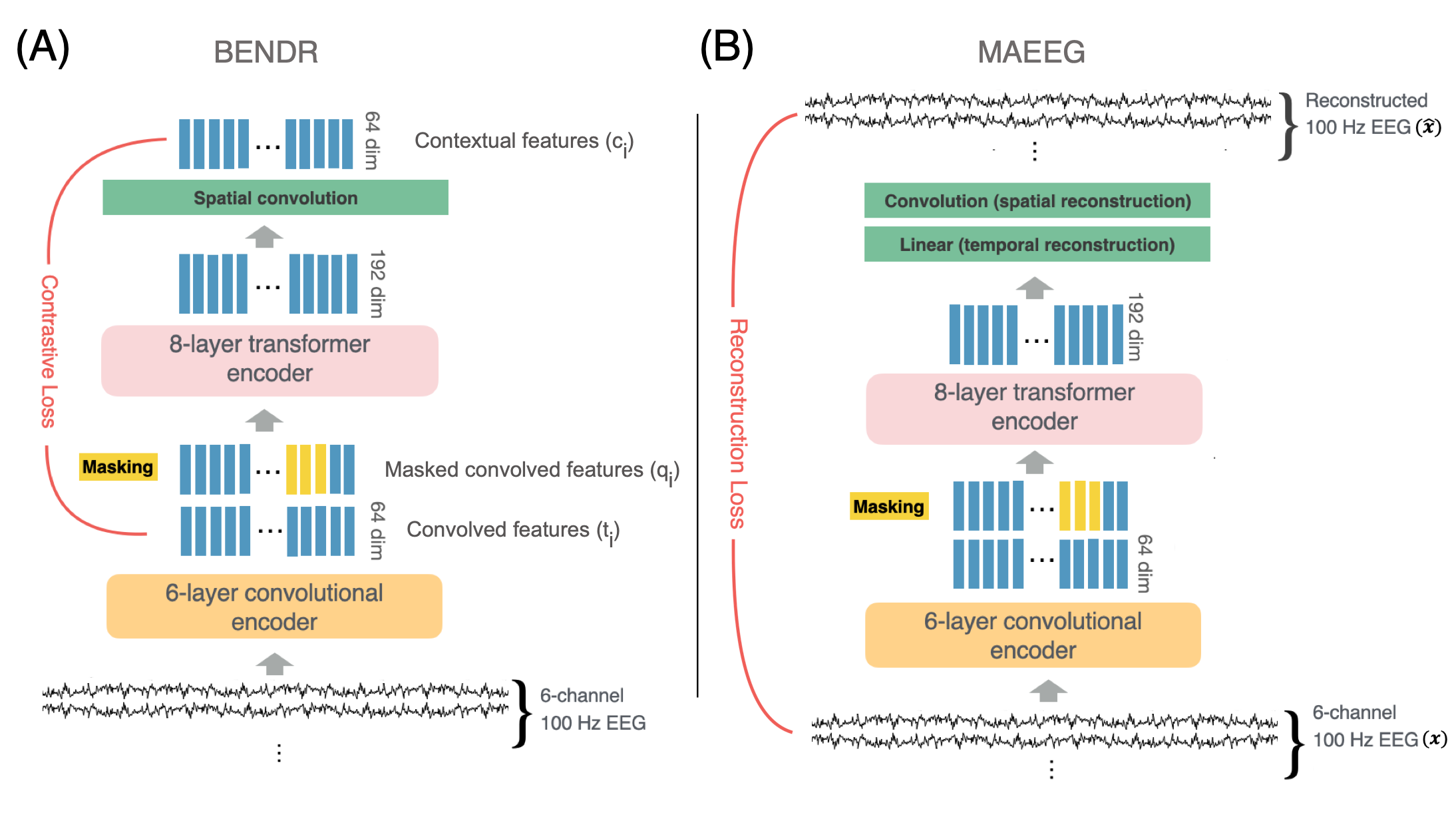}}
\caption{\textbf{Comparison between two reconstruction-based SSL model architecture.} \textbf{A.} BENDR learns representations through contrastive loss between convolved and contextual features, while \textbf{B.} MAEEG learns representations through reconstruction loss between input and output EEG signals.}
\label{fig:SSL_models}
\end{figure*}

\paragraph{BENDR.} The model architecture of BENDR is shown in Figure \ref{fig:SSL_models}A. First, the 6-channel 100Hz raw EEG signals are the input to the 6 convolutional layers, which encode the raw EEG into 64-dimensional convolved features ($t_i$, $\sim$1.05Hz). Dropout and GroupNorm are added to each convolutional layer and GELU is the output activation function. A mask $I_m$ is generated to ``mask out" part of the features $t_i$ where $i \in I_m$ (i.e., set the activation to Gaussian noise $M\sim \mathcal{N}(0,\,\sqrt{1/64})$). The masked features $q_i$ is then sent to an 8-layer transformer encoder and encoded as 192-dimensional output features. The positions are encoded using a convolutional layer, which acts as relative positional embedding. Finally, a final convolutional layer maps the features back to 64-dimensional contextual features $c_i$. Contrastive loss is calculated by comparing $c_i$ with the unmasked features $t_i$.

\paragraph{MAEEG.} The MAEEG model has a similar architecture to BENDR (Figure~\ref{fig:SSL_models}B), but has two additional layers to map the transformer output back to the raw EEG dimensions. A reconstruction loss is calculated by comparing the reconstructed EEG ($\bf \hat{x}$) and input EEG ($\bf x$) signals as $1-\frac{{\bf \hat{x}}\cdot{\bf x}}{\|{\bf \hat{x}}\| \|\bf {x}\|}$ (Figure \ref{fig:SSL_models} B). The key difference between BENDR and MAEEG is that instead of using contrastive learning, MAEEG learns representations by minimizing the reconstruction loss (Appendix \ref{sec:maeeg_reconstruction}).
\subsection{Features for Downstream classification}
After SSL pretraining, we examined downstream task performances by adding a linear classification layer on top of either of the two features: the convolved features, $t_i$, which are the output of the convolutional encoder, and the contextual features, $c_i$, which are the output of the transformer encoder. To build downstream classification model for classifying sleep stages, we preserve most of the SSL architecture with the pretrained encoders, and simply add a fully-connected classification layer which takes the SSL features as input and generates probabilities for predicting the 5 sleep stages (Figure \ref{fig:DS_models}). The choice of adding a linear classifier instead of a non-linear one is to prevent the classifier from being too powerful and making it difficult to evaluate the representations learned from SSL pretraining.  To better understand representations learned from different stages of training (i.e.,\ SSL pretraining with no labels and downstream supervised training with labels), we conducted two downstream classification analyses: (1) To examine low-level features $t_i$ learned solely from the SSL pre-training, we use $t_i$ while ``freezing" the encoder (meaning that the weights of the encoder are fixed) during training for the downstream classification. Only the classification layer gets trained in this condition (Figure \ref{fig:DS_models} A); (2) To examine the pre-trained model as an initialization for supervised training, we use $c_i$ while fine-tuning the encoders and the classification layer during downstream classification (Figure \ref{fig:DS_models} B). Specifically, we extracted $c_i$ from layer 2 as the SSL features to conduct all the analyses given that it showed the best results among all the transformer layers. We conducted these analyses on MAEEG, BENDR and a baseline supervised model which has the same architecture as other models but the encoders are randomly initialized, i.e., \ no SSL pretraining prior to the downstream classification task.

For each downstream task, we trained 100 epochs in total and validated the model every 2 epochs. We selected the models with highest validation accuracy to run the test set and get the final accuracy shown in Figure \ref{fig:ssl_performance}. We also investigated how masking in reconstruction-based SSL could affect EEG representation learning by modifying the masking probability and lengths (See Appendix \ref{sec:mask_effect}). We applied analyses with various number of subjects or percent of sessions to examine model performance given different amount of labels. 
\label{sec:features_downstream}

\section{Results}

\subsection{MAEEG Learned Useful Representations for Classifying Sleep Stages}
\begin{figure*}[ht]
\centerline{\includegraphics[width=\columnwidth]{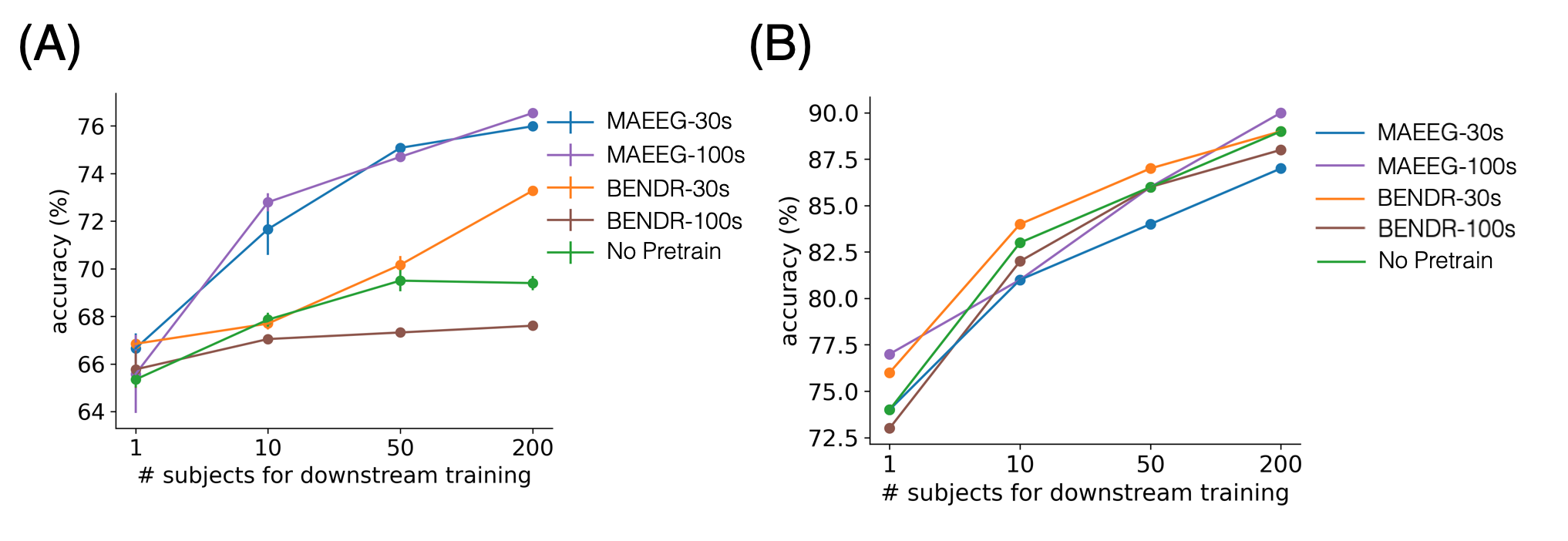}}
\caption{\textbf{Model comparison on classifying sleep stages} \textbf{A.} Classification performance with different numbers of downstream subjects using convolved features when the encoders were frozen. \textbf{B.} Classification performance using contextual features when the encoders were fine-tuned. Chance level was 20\%. The error bars indicate the standard deviation of results from 3 training samples.}
\label{fig:ssl_performance}
\end{figure*}

We first examine convolved features $t_i$ learned solely by SSL pretraining by freezing the convolutional encoders during training for the sleep stage classification task. We found that representations pretrained by MAEEG (both 30s and 100s samples input) significantly outperformed the representations trained by BENDR and the supervised baseline model across different amount of subjects for training (Figure \ref{fig:ssl_performance}A). This suggests that representations learned from MAEEG pretraining do capture features that are useful for classifying sleeping stages.   

Next, to examine how SSL pretraining may improve the supervised learning performance, we used contextual features $c_i$ and fine-tuned the model during the sleep stage classification task. We found that when only given one subject's labels (containing all sleep stages), MAEEG pretrained on 100s (MAEEG-100s) outperformed other models, while BENDR pretrained on 30s (BENDR-30s) performed the best on 10 subjects' and 50 subjects' labels (Figure\ref{fig:ssl_performance}B). The best classification performance (90\% accuracy) was from MAEEG-100s. Interestingly, MAEEG-100s performed significantly better compared to MAEEG-30s, and an opposite effect on BENDR was observed - BENDR-30s performed significantly better compared to BENDR-100s, suggesting that representations learned from the two reconstruction-based SSL models are distinct and the sample length for pretraining has a big effect on downstream performance. To understand how the pretraining differs under different experimental settings, we further examined the attention maps learned during pretraining. We found that models that learned to attend to distant features (e.g.\ 20 seconds away) performed better than models that learned to attend neighboring features (See Appendix \ref{sec:attention}, Figure\ref{sec:attention}). Also, the advantage of MAEEG-pretrained convolved features did not reflect in performance when the encoders were fine-tuned, suggesting that while MAEEG may learn some useful features to discriminate sleep stages through SSL pretraining, but may also hinder the models' ability to learn at the same time.

\subsection{Masking Effects in MAEEG}
Next, we examined how different ways of masking during reconstruction-based SSL pretraining may affect representation learning and downstream classification performance using contextual feature $c_i$ with finetuning, by varying the mask rate and the number of mask chunks (See Appendix \ref{sec:mask_effect}). We first examined the SSL loss during pretraining, and found that higher percentage of masking with less chunks (which yields longer masks) yielded higher SSL loss (i.e.\, more difficult to train). We then examined the downstream model performance and found that the model performances differ most when given the least number of labels - such as 0.1\% (i.e.\ $\sim$4 hours of labels) of downstream training data (Figure \ref{fig:mask_result}B). With such small number of labels, we found that masking 75\% of the tokens during SSL pretraining performed better overall compared to masking 50\% and 25\% of the tokens, and models with a single chunk performed better than models with multiple chunks. All models with MAEEG pretraining performed better than the models without pretraining (Figure \ref{fig:mask_result}C). Finally, we examined whether such a trend (single chunk with longer mask performs better) preserved when we vary the mask span of a single mask chunk, and whether it generalizes to BENDR. We found a consistent pattern across both MAEEG and BENDR, that models pretrained with longer mask span performs better than models pretrained with shorter mask span when there only a small number of labels (Figure \ref{fig:mask_result}D). This finding is likely due to the fact that a single and long mask during pretraining forced the model to learn a more difficult task and yield more useful representations for sleep stage classification.

\section{Discussion and Limitations}
In the current study, we present MAEEG and compare it with another SSL model, BENDR, to examine how reconstruction-based SSL could be applied to learn meaningful EEG representations. Specifically, we examined how the representations learned from these two SSL approaches could be applied to different downstream conditions by varying the amount of downstream training data. To compare model performance while controlling the subject variance, we showed the results in which we sampled data from different number of subjects (Figure \ref{fig:ssl_performance}). While we showed that representations learned from MAEEG yielded better performance, we could not exclude the possibility that some of the effect might still come from the specific subjects that we sampled from. We also conducted analyses by sampling different sessions from every subject as shown in the masking analyses (Figure \ref{fig:mask_result}). However, we are still sensitive to potential label imbalance and to the large subject variance in a small amount of data which could make the training more difficult. Subject variance and imbalanced classification are common challenges for learning from bio-signal data. Future studies should take these issues into account while examining representation learning for EEG or bio-signals.  

One interesting observation in the current study is that masking out more signal (e.g.,\ 75\%) in the latent space yielded better downstream performance compared to masking out less signal (e.g.,\ 25\%). While this result may not be intuitive, the same trend was observed in computer vision where masking more of the signal during SSL pretraining could result in better downstream performance \citep{he2022masked}. The reason behind this could be that asking the model to reconstruct a larger amount of missing samples forces the model to learn the relationship between the remaining context and the missing part of the signal, instead of just interpolating it. This trend can be dependent on the downstream task. We suggest future studies to always evaluate the SSL model based on its downstream performance instead of pretraining performance. 

Finally, the computational cost for MAEEG and BENDR are similar in the current study, given that both models have the signal masked in the latent space. Such choice of masking might be the reason why MAEEG only learns to reconstruct slow signal variations (Section \ref{sec:maeeg_reconstruction}). Masking in the raw signal space may help MAEEG to learn to reconstruct more fine-grained signals but could also make the computation more expensive. Alternatively, masking in the time-frequency spectrum domain could be beneficial for EEG representation learning \citep{xu2022masked}.

\section{Conclusion}
Representation learning for EEG can be useful given the difficulty in collecting large amount of data and labels. In this study, we proposed a reconstruction-based SSL model, MAEEG, for learning EEG representations by reconstructing EEG signals using masked auto-encoder architecture. We tested our model on sleep stage classification and found that MAEEG can learn useful representations solely from the pre-text task which boost the downstream classification performance. We also found that different ways of masking and sample lengths selected during SSL pre-training can significantly affect the downstream classification performance. We encourage future studies to examine how reconstruction-based SSL may help representation learning on other time-series data for health.

\bibliographystyle{plainnat}
\bibliography{references}

\newpage

\appendix
\renewcommand{\thefigure}{A.\arabic{figure}}
\setcounter{figure}{0}

\section{Appendix}

\subsection{EEG Signal Reconstruction}
Note that the current MAEEG architecture (Figure \ref{fig:SSL_models}B) is inspired by the MAE methods proposed in \citet{he2022masked} rather than adapted the model architecture. Instead of using transformer layers as decoder as in \citet{he2022masked}, we simply add two convolution layers in MAEEG to allow the network to reconstruct the raw EEG signal temporally and spatially. By visualizing the reconstructed signal, we observed that MAEEG mainly learned to reconstruct the mean of the short-timescale signals, and capture the long-timescale fluctuations. Based on the results, capturing long-timescale fluctuations seems already beneficial for learning representations for sleep stage classification, possibly due to the fact that such specific task relies more on long-timescale information. It is likely that other downstream tasks would require the model to learn representations by reconstructing more fine-grained information.
\label{sec:maeeg_reconstruction}

\subsection{Masking Effect Analyses}
\begin{figure*}[ht]
\centerline{\includegraphics[width=\columnwidth]{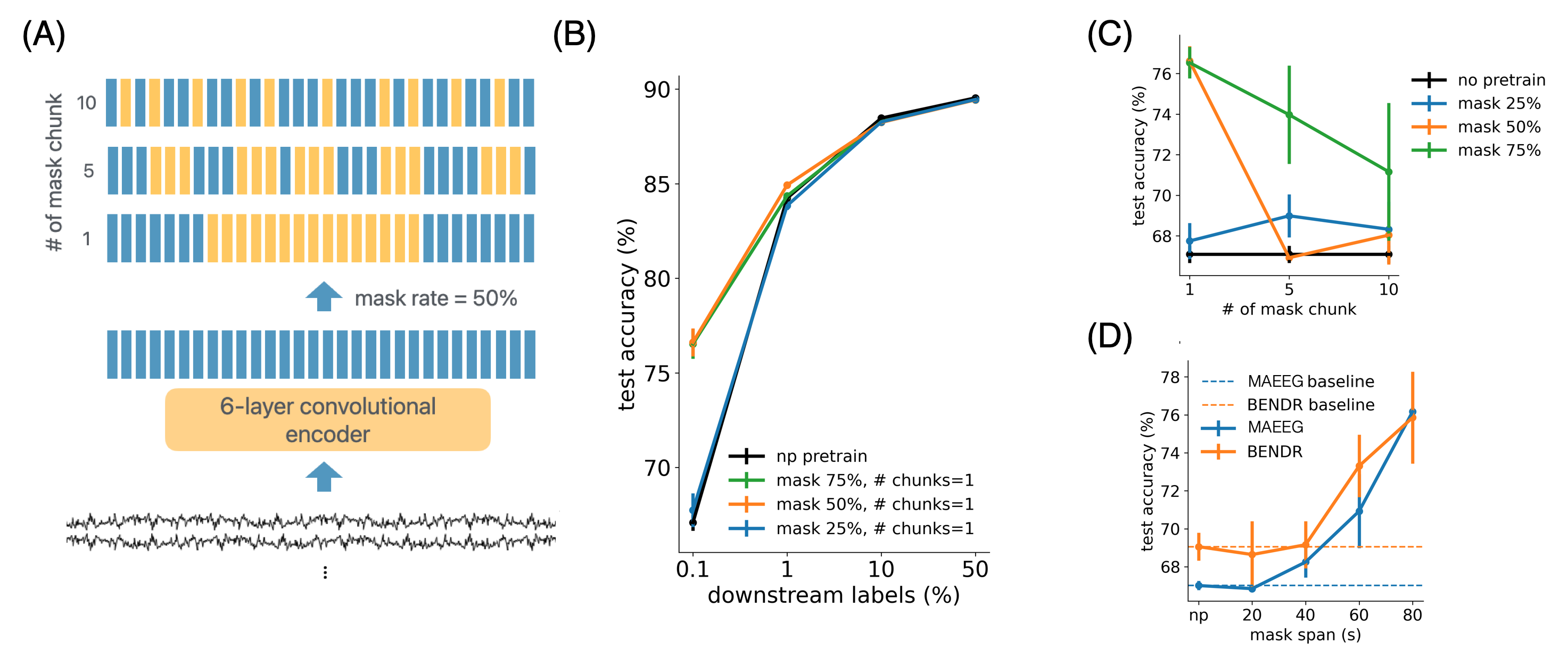}}
\caption{\textbf{Masking effect on classification performance} \textbf{A.} Example of generating different masks for SSL pretraining by varying the mask rate and number of mask chunks. \textbf{B.} Classification performance using fine-tuned contextual features with various masking conditions and percentage of training sessions. \textbf{C.} Classification results using 0.1\% of labels with different masking conditions. Overall, we found that masking 75\% of the signals with 1 chunk results in the best performance. \textbf{D.} Classification results using 0.1\% of labels when varying the masking span. We found that longer mask span yields better results for both MAEEG-100s and BENDR-100s. np = no pretrain}
\label{fig:mask_result}
\end{figure*}

The masking during SSL pretraining determines the difficulty of the SSL task, and the representations learned from such task. Therefore, it is critical to understand what could be an ideal way of masking for learning EEG signals. In the original BENDR, they set a probability (p=0.065) for each token to be the beginning of a mask with a length of 10 tokens. The problem with this is that for each sequence, the mask can vary a lot - sometimes most of the tokens can be masked with overlap, and sometimes there is no token being masked - which may cause the instability of representation learning. Here, we tried to use a more systematic way to generate the mask, by determining the mask rate and the number of mask chunks. These two factors would determine how long each mask looks like, as illustrated in Figure \ref{fig:mask_result} A. We did two analyses: 
\begin{enumerate}
    \item  Pretrained the models with 3 mask rates: 75\%, 50\% and 25\%, and 3 numbers of mask chunks: 1, 5, and 10.
    \item Pretrained models with single mask chunk but vary the mask span to further confirm the effects.
\end{enumerate}
The resulting pretrained models were evaluated by the downstream performance when using the learned features for classification.
We conducted all the analyses on BENDR and MAEEG pretrained on 100s samples because models pretrained on 30s samples did not have enough tokens for examining the effect of these mask variations. Also, we examined varying percentage of training samples during the downstream task to examine the effect of masking under different sample size conditions. Note that instead of varying the number of subjects as in the main task, we varied the percentage of data sampled from each subject, which eliminates the potential bias coming from the specific sampled subjects.   
\label{sec:mask_effect}


\subsection{Attention Learned from SSL}
\begin{figure*}[ht]
\centerline{\includegraphics[width=\columnwidth]{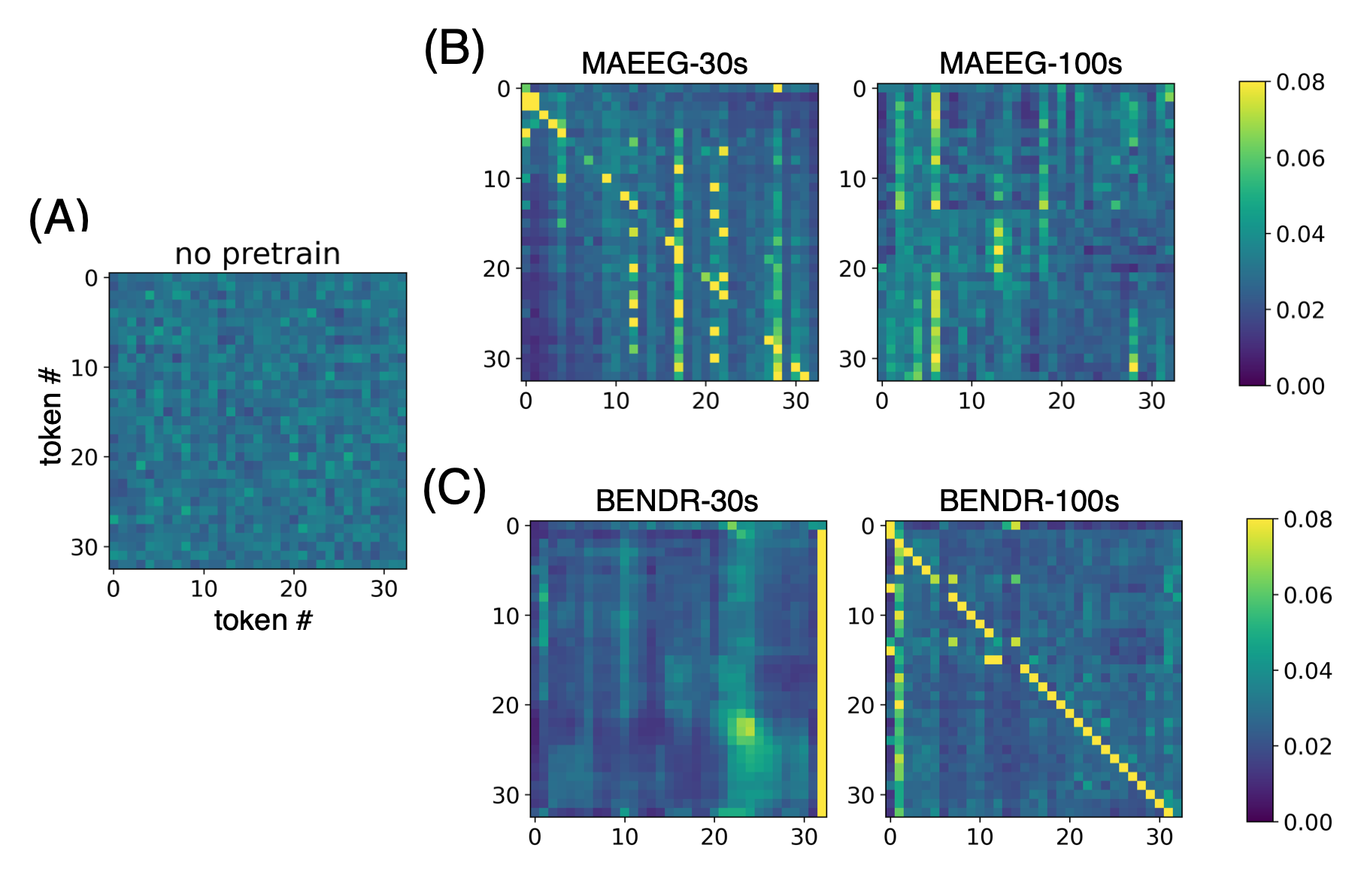}}
\caption{\textbf{Visualizing attention after pretraining} \textbf{A.} Initialized attention. \textbf{B.} (left) attention learned from MAEEG-30s pretraining; (right) attention learned from MAEEG-100s pretraining \textbf{C.} (left) attention learned from BENDR-30s pretraining; (right) attention learned from BENDR-100s pretraining.}
\label{fig:attention}
\end{figure*}

To understand how the EEG representation learned from reconstruction-based SSL help sleep classification, we visualized the attention from the transformer layer where features were extracted (i.e.\ layer 2) after SSL pretraining but before extensive downstream supervised training (we slightly tuned the models with 1 epoch with 1 subject's data just to get stable representations.) We extract the token-by-token attention maps for the baseline supervised model and the models pretrained by MAEEG and BENDR. We visualized the attention maps from a sample where the label is N2 (Figure \ref{fig:attention}). Compared to the attention from baseline models which is equally distributed as shown in Figure \ref{fig:attention} A, MAEEG- and BENDR- pretrained models showed distinct attention patterns learned from pretraining. Interestingly, we observed that MAEEG-100s and MAEEG-30s actually learned very different attention patterns, with MAEEG-30s learning to attend locally to neighboring tokens while MAEEG-100s learning to attend more distant tokens (Figure \ref{fig:attention} B). On the other hand, BENDR-100s and BENDR-30s also learned very different attention patterns, with BENDR-100s learning to attend locally and BENDR-30s learning a more global distributed attention. This suggests that representation learning can be very different based not only on the model architecture, but also on subtle SSL settings such as the length of pretraining samples. Moreover, MAEEG-30s and BENDR-100s performed worse on the classification task compared to their counterparts (Figure \ref{fig:ssl_performance}B), suggesting that local attention acquired from pretraining may not be desirable for the sleep stage classification task, possibly due to the fact that this specific task relies on long-timescale information.
\label{sec:attention}

\newpage

\begin{figure*}[ht]
\centerline{\includegraphics[width=\columnwidth]{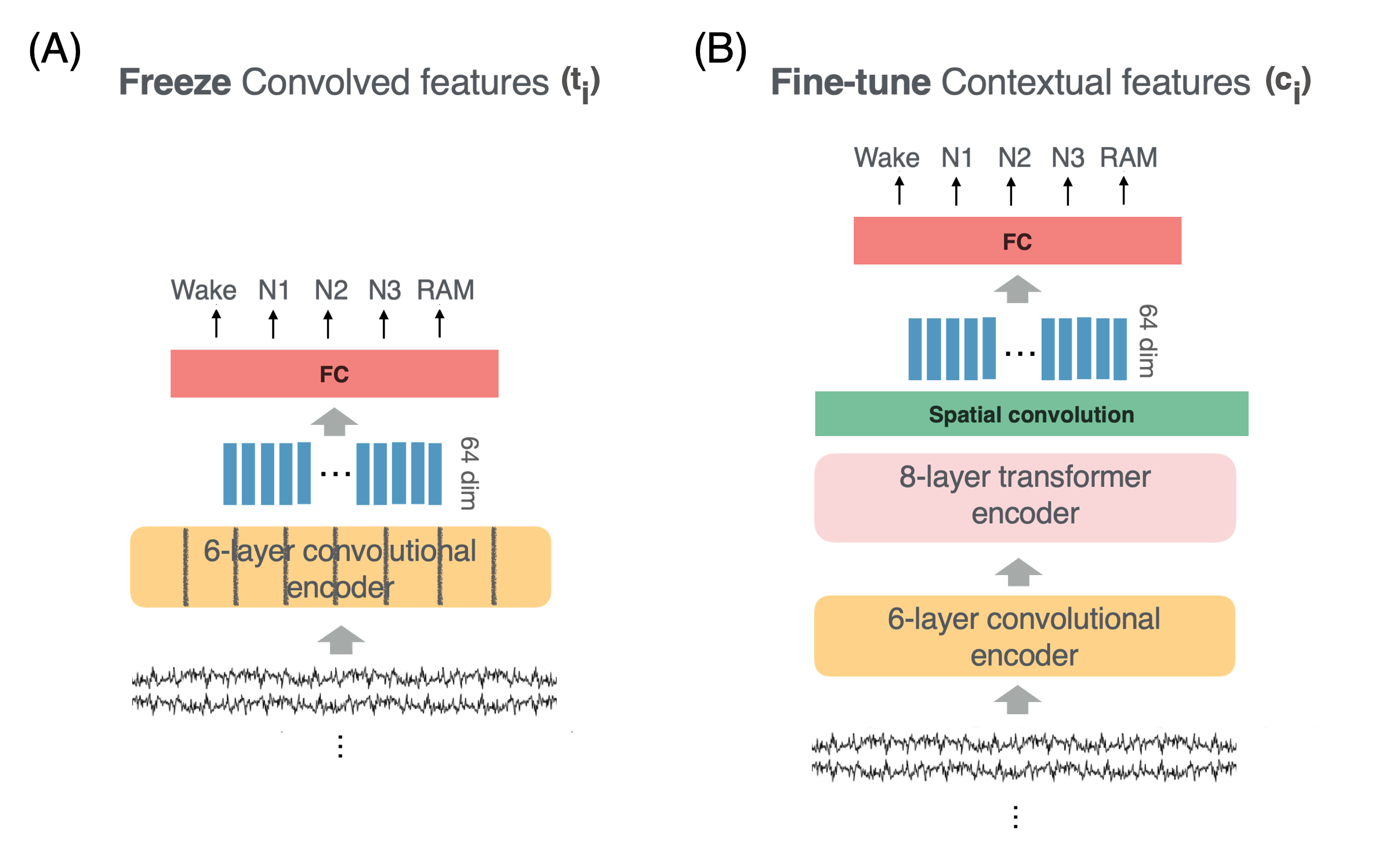}}
\caption{\textbf{Downstream model architecture} \textbf{A.} Classification using convolved features $t_i$ with encoder frozen to examine the representation learned solely by SSL without labels.  \textbf{B.} Classification using contextual features $c_i$. The encoders are fine-tuned during downstream training to examine the pre-trained model as an initialization for supervised training.}
\label{fig:DS_models}
\end{figure*}

\end{document}